\pdfoutput=1
\documentclass{sig-alternate-10pt}
\usepackage[margin=1in]{geometry}
\usepackage[T1]{fontenc}

\usepackage[utf8]{inputenc} 
\usepackage{amsmath}
\usepackage{mathtools}
\usepackage{graphicx} %
\usepackage{url}
\usepackage{enumitem} %

\usepackage{doi}
 \usepackage{bm} %
\usepackage{stmaryrd} %

\usepackage{upgreek} %
\usepackage{bbding} %

\usepackage{listings} %

\usepackage{tabularx} %
\usepackage{booktabs} %
\usepackage{multirow}
\usepackage{hhline} %
\usepackage{diagbox}
\usepackage{pgfplotstable} %

\usepackage{xcolor,colortbl}    %

\usepackage{tikz} %
\usetikzlibrary{matrix}
\usetikzlibrary{calc}
\usetikzlibrary{math}
\usetikzlibrary{arrows.meta,positioning}
\usetikzlibrary{intersections, pgfplots.fillbetween}

\usetikzlibrary{positioning}
\newcommand*\circled[1]{\tikz[baseline=(char.base)]{
            \node[shape=circle,fill=black,draw,text=white,inner sep=1.2pt] (char) {#1};}}

\usepackage{easybmat} %

\usepackage{adjustbox}
\def\eox{\unskip\kern 10pt{\unitlength1pt\linethickness{.4pt}$\diamondsuit${}}} %

\usepackage{rotating}		%

\usepackage{xspace} %

\usepackage{subcaption} %
\newcommand{\change}[1]{{\leavevmode\color{blue}#1}}

\newcommand{\revchange}[1]{{\leavevmode#1}}

\newcommand{\hide}[1]{} 
\newcommand{\hidetwo}[2]{}

\usepackage[ruled,noend,linesnumbered]{algorithm2e} %

\DontPrintSemicolon     %
\SetNlSty{}{}{}                %
\SetAlgoInsideSkip{smallskip}   %

\SetAlFnt{\small}			%
\SetAlCapFnt{\small}		%
\SetAlCapNameFnt{\small}

\usepackage{hyperref}   %
\hypersetup{                                                            %
	pdfpagemode=UseNone,               %
    pdfstartview=Fit,
    pdfpagelayout=SinglePage,       %
    bookmarks=false,
    bookmarksnumbered=true,
    bookmarksopen=false,             %
    pdftex=true,
    unicode=true,
    colorlinks=true,       %
    linkcolor=blue,          %
    citecolor=cyan,  %
    filecolor=magenta,      %
     urlcolor=blue           %
}

\usepackage[capitalise,nameinlink]{cleveref} %

\usepackage{aliascnt}  	

\newtheorem{theorem}{Theorem} %

\newaliascnt{corollary}{theorem}

\aliascntresetthe{corollary}

\newaliascnt{example}{theorem}

\aliascntresetthe{example}

\newaliascnt{definition}{theorem}
\newtheorem{definition}[definition]{Definition}
\aliascntresetthe{definition}

\newaliascnt{proposition}{theorem}

\aliascntresetthe{proposition}

\newaliascnt{lemma}{theorem}

\aliascntresetthe{lemma}

\newaliascnt{conjecture}{theorem}

\aliascntresetthe{conjecture}

\newtheorem{questionW}{Question}
\newtheorem{resultW}{Result}

{\end{itshape}
}
\usepackage{tcolorbox}		%
\tcbuselibrary{breakable,skins}		%

\tcbset{examplestyle/.style={
		enhanced jigsaw,	%
		colback=blue!08,	%
		colframe=blue!08,	%
		arc=3mm,
		boxrule=0pt,		%
		left=1mm,
		right=1mm,
		left skip= 0mm,  %
		right skip= 0mm, %
		top=1mm,		%
		bottom=1mm,		%
		breakable,		%
		parbox = false,		%
		before={\par\pagebreak[0]\vspace{1mm}\parindent=0pt},		
		after={\par\pagebreak[0]\vspace{1mm}\parindent=0pt},				
		bottomrule = 0mm,
		boxsep = 0mm,					%
		topsep at break=0pt,			%
		bottomsep at break=0pt,			%
		pad at break=0mm,
		pad before break=1mm,		
		pad after break=1mm,		
		bottomrule at break=0mm,
		toprule at break=0mm,		
		}}

\usepackage{footnote}

\tcolorboxenvironment{example}{examplestyle}
\makeatletter
\DeclareRobustCommand*\uell{\mathpalette\@uell\relax}
\newcommand*\@uell[2]{
  \setbox0=\hbox{$#1\ell$}
  \setbox1=\hbox{\rotatebox{10}{$#1\ell$}}
  \dimen0=\wd0 \advance\dimen0 by -\wd1 \divide\dimen0 by 2
  \mathord{\lower 0.1ex \hbox{\kern\dimen0\unhbox1\kern\dimen0}}
}
\makeatother

\usepackage{marginnote}

\newcommand{\introparagraph}[1]{\textbf{#1.}} %

\renewcommand{\epsilon}{\varepsilon} %

\newcommand{\datarule}{{\,:\!\!-\,}} %
\renewcommand{\vec}[1]{\boldsymbol{\mathbf{#1}}}

\newcommand{\N}{\mathbb{N}} %
\newcommand{\R}{{\mathbb{R}}} %
\renewcommand{\O}{{\mathcal{O}}} %
\renewcommand{\P}{{\mathbb{P}}} %
\newcommand{\E}{\mathbb{E}}

\usepackage{scalerel}
\usepackage{tikz}
\usetikzlibrary{svg.path}

\definecolor{orcidlogocol}{HTML}{A6CE39}
\tikzset{
  orcidlogo/.pic={
    \fill[orcidlogocol] svg{M256,128c0,70.7-57.3,128-128,128C57.3,256,0,198.7,0,128C0,57.3,57.3,0,128,0C198.7,0,256,57.3,256,128z};
    \fill[white] svg{M86.3,186.2H70.9V79.1h15.4v48.4V186.2z}
                 svg{M108.9,79.1h41.6c39.6,0,57,28.3,57,53.6c0,27.5-21.5,53.6-56.8,53.6h-41.8V79.1z M124.3,172.4h24.5c34.9,0,42.9-26.5,42.9-39.7c0-21.5-13.7-39.7-43.7-39.7h-23.7V172.4z}
                 svg{M88.7,56.8c0,5.5-4.5,10.1-10.1,10.1c-5.6,0-10.1-4.6-10.1-10.1c0-5.6,4.5-10.1,10.1-10.1C84.2,46.7,88.7,51.3,88.7,56.8z};
  }
}

\RequirePackage{etoolbox}
\DeclareRobustCommand\orcidicon[1]{\href{https://orcid.org/#1}{\mbox{\scalerel*{
\begin{tikzpicture}[yscale=-1,transform shape]
\pic{orcidlogo};
\end{tikzpicture}
}{|}}}}

\def\polylog{\operatorname{polylog}} %

\newcommand{\sparseBMM}{\textsc{SparseBMM}}
\newcommand{\hyperclique}{\textsc{Hyperclique}}

\newcommand{\dom}{{\mathtt{dom}}}

\newcommand{\PSQL}{\textsc{PSQL}\xspace}

\newcommand{\BATCH}{\textsc{JoinFirst}\xspace}

\newcommand{\ANYKPART}{\textsc{anyK-part}\xspace}

\usepackage{booktabs} %
\graphicspath{{./figs/}}
\usepackage{numprint}
\usepackage{multirow}
\usepackage{makecell}

\usepackage{balance}
\usepackage{tikz}
\usepackage{tikz,tikz-qtree,tikz-qtree-compat,tikz-3dplot}
\usetikzlibrary{shapes,decorations.markings,arrows.meta,fit}

\definecolor{dkgreen}{rgb}{0,0.6,0}
\newcommand{\cc}{olive}
\newcommand{\algocomment}[1]{\textcolor{\cc}{{//#1}}}

\newcommand{\TT}{\ensuremath{\mathrm{TT}}}

\renewcommand{\H}{\mathcal{H}}
\newcommand{\G}{\mathcal{G}}

\newcommand{\subw}{\textsf{subw}\xspace}

\makeatletter
\newcommand{\markZwicky}[1][]{\pgfutil@ifnextchar({\mark@Zwicky{#1}}{\mark@Zwicky{#1}()}}
\def\mark@Zwicky#1(#2)#3{%
   \tikz[every Zwicky picture,#1]{%
     \node[every Zwicky node,draw=none,inner sep=+\z@,outer sep=+\z@] {#3};
     \def\tikz@Mark@name{#2}%
     \ifx\tikz@Mark@name\pgfutil@empty\else
       \tikzset{every Zwicky node/.append style={name={#2}}}%
     \fi
     \node[every Zwicky node,overlay] {\phantom{#3}};
   }%
}
\newcommand{\tikzZwicky}[1][]{%
  \def\tikz@Zwicky@args{#1}%
  \let\tikz@Zwicky@list\pgfutil@gobble
  \let\tikz@Zwicky@first\pgfutil@empty
  \pgfutil@ifnextchar(\tikz@Zwicky@collect\tikz@Zwicky@finish
}
\def\tikz@Zwicky@collect(#1){%
  \ifx\tikz@Zwicky@first\pgfutil@empty
    \edef\tikz@Zwicky@first{#1}%
  \else
    \edef\tikz@Zwicky@list{\tikz@Zwicky@list,#1}%
  \fi
  \pgfutil@ifnextchar(\tikz@Zwicky@collect\tikz@Zwicky@finish
}
\def\tikz@Zwicky@finish{%
  \tikz[remember picture,overlay]
    \draw[every Zwicky connector,/expanded=\tikz@Zwicky@args]
      (\tikz@Zwicky@first) [/expanded={@Zwicky@list/.list={\tikz@Zwicky@list}}] [every Zwicky connect finish/.try];
}
\pgfkeys{/expanded/.code={\edef\pgfkeys@temp{{#1}}\expandafter\pgfkeysalso\pgfkeys@temp}}
\makeatother
\tikzset{
  @Zwicky@list/.style={insert path={to[every Zwicky connector how/.try] (#1)}},
  every Zwicky picture/.style={
    baseline,
    remember picture,
  },
  every Zwicky node/.style={
    remember picture,
    anchor=base,
    inner sep=+2pt
  },
  every Zwicky connector/.style={
    ultra thick,
    red!80!black,
    draw opacity=.5,
    line cap=round,
    line join=round
  }
}

\newtheorem{remark}{Remark}

\makeatletter
\let\oldnl\nl%
\newcommand{\nonl}{\renewcommand{\nl}{\let\nl\oldnl}}%
\makeatother

\makeatletter
\newcommand{\algocf@VslineMine}[1]{%
  \par\nointerlineskip%
  \algocf@push{\skiprule}%
  \hbox{\vrule%
    \vtop{\algocf@push{\skiptext}%
      \vtop{\algocf@addskiptotal\advance\hsize by -\skiplength #1}}}%
  \algocf@pop{\skiprule}}%
\newcommand\Block[1]{%
    \algocf@VslineMine{#1}%
}
\makeatother

\def\rel#1{\mathsf{#1}}
\def\att#1{\mathrm{#1}}

\newcommand{\Otilde}{\tilde{\O}}

\newcommand{\LexArr}{\parbox{.3cm}{\tikz{\draw[->](0,0)--(.25cm,0);}}}
\newcommand{\calT}{\mathcal{T}}
\newcommand{\calS}{\mathcal{S}}
\newcommand{\calP}{\mathcal{P}}

\newcommand{\joinindex}[2]{\ensuremath{\textrm{JoinIndex}_{#1 \rightarrow #2}}}
\newcommand{\relind}{\texttt{rel}}
\newcommand{\tupval}{\texttt{val}}

\newcommand{\parent}{\texttt{pr}}
\newcommand{\sumrank}{\mathcal{W}}
\newcommand{\lexrank}{L}

\newcommand{\ar}{\texttt{ar}}

\newcommand{\SemiJoin}{\ensuremath{\ltimes}}
\newcommand{\True}{\ensuremath{\texttt{True}}}
\newcommand{\False}{\ensuremath{\texttt{False}}}
\newcommand{\tupopt}{\ensuremath{\texttt{opt}}}

\newcommand{\wt}{\ensuremath{\texttt{w}}}
\newcommand{\prio}{\ensuremath{\texttt{prio}}}

\usepackage{etoolbox} %

\makeatletter
\patchcmd{\algocf@makecaption@ruled}{\hsize}{\linewidth}{}{} %
\patchcmd{\@algocf@start}{-1.5em}{0em}{}{} %
\makeatother

\SetAlCapHSkip{0pt} %
\usepackage[style=trad-abbrv,backend=bibtex,doi=true,isbn=false,url=true,eprint=false,maxbibnames=99]{biblatex} 
\usepackage{titlesec}

\def\email#1{{{\affaddr{\par #1}}}}       %

\begin{document}

\title{Ranked Enumeration for Database Queries}

\numberofauthors{3}
\author{
Nikolaos Tziavelis\titlenote{Work done while at Northeastern University.}\\
\affaddr{UC Santa Cruz}\\
\email{ntziavel@ucsc.edu}
\and 
Wolfgang Gatterbauer\\
\affaddr{Northeastern University}\\
\email{w.gatterbauer@northeastern.edu}
\and 
Mirek Riedewald\\
\affaddr{Northeastern University}\\
\email{m.riedewald@northeastern.edu}
}

\maketitle

\begin{abstract} 
Ranked enumeration is a query-answering paradigm where 
the query answers are returned incrementally in order of importance
(instead of returning all answers at once).
Importance
is defined 
by a ranking function that can be specific to 
the application,
but typically involves either
a lexicographic order (e.g., ``ORDER BY R.A, S.B'' in SQL) 
or a weighted sum of attributes (e.g., ``ORDER BY 3*R.A + 2*S.B'').
\revchange{Recent work has}
introduced \emph{any-$k$ algorithms} 
for (multi-way) join
queries, which 
push ranking into joins 
and avoid materializing
intermediate results
until necessary. 
The top-ranked answers are returned asymptotically faster
than the common join-then-rank approach of
database systems,
resulting in orders-of-magnitude speedup in practice.

In addition to their practical usefulness, 
\revchange{these techniques}
complement a long line
of theoretical research on \emph{unranked enumeration},
where answers are also returned incrementally, but with no explicit ordering requirement.
For a broad class of ranking functions with certain monotonicity properties,
including lexicographic orders and sum-based rankings,
the ordering requirement surprisingly does not increase the asymptotic time or space
complexity, apart from logarithmic factors.

\revchange{A key insight
is the connection between ranked enumeration for database queries and
the fundamental task of computing the $k^\textrm{th}$-shortest path in a graph.
Although this connection is important for grounding the problem in the literature,
it can obfuscate the simplicity of the algorithm.}
In this article, we adopt a pragmatic approach and present a slightly simplified version
of the algorithm without the shortest-path interpretation.
We believe that this will benefit practitioners looking to implement and optimize any-$k$
approaches.
\end{abstract}

\begin{figure}
    \centering
    \includegraphics[width=0.95\linewidth]{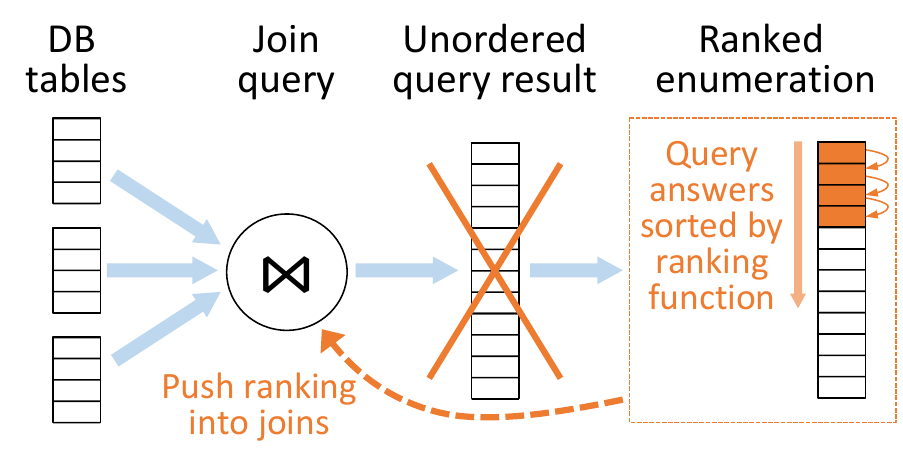}
    \caption{Enumerating the query answers in ranked order without first materializing the unordered query result.
    \revchange{\emph{Sorting is pushed into
    the join operation}}
    so that joining and ranking are interleaved.
    }
    \label{fig:pushdown}
\end{figure}

\section{Introduction}
\label{sec:intro}

Data analytics queries can generate large intermediate or final results,
rendering
data systems unresponsive.
A primary culprit is the join operator, which combines data from different tables,
potentially causing a combinatorial explosion in the output.
Consequently, traditional join-processing techniques 
can
become infeasible,
or they simply take too long before delivering
any answer to the user or to the next step in a data-processing pipeline.
Work on \emph{enumeration}~\cite{bagan07constenum,DBLP:journals/sigmod/Segoufin15}
addresses this by returning query answers incrementally as quickly as possible,
even when the full query output is too large to compute.
However, enumeration traditionally does not support a desired order (or \emph{ranking}) specifying,
which answers should be returned first.
We thus refer to it as \emph{unranked enumeration}.
In practice, certain answers may be preferred over others based on
some notion of importance or relevance. For instance, higher importance may
be assigned to newer or more trusted data. 
\emph{Ranked enumeration}~\cite{deep21,tziavelis20vldb} 
therefore augments enumeration with a total-order feature over the query answers,
formalized by a ranking function (e.g., expressed by an ORDER BY clause in SQL).

Database systems today follow a join-then-rank approach, i.e.\
they first compute \emph{all join answers} and then apply the ranking (by sorting either incrementally or in batch).
One way to think about the improvement we seek is that we want to ``push'' the ranking operator deeper into the query plan.
While this resembles typical database optimizations, such as pushing
projections before joins, the task is more challenging, because
join and ranking operators generally do not commute. 
Novel algorithms are required,
where joining
and ranking are interleaved.\footnote{Even simpler top-1 queries 
are not efficiently supported by current systems.
For a minimum example in PostgreSQL, see slide 20: \url{https://northeastern-datalab.github.io/cs7240/sp24/download/cs7240-T3-U1-Acyclic_Queries.pdf}.}

\introparagraph{Performance Goal}
How can performance for such an algorithm be measured?
The top-ranked answers should be returned quickly
without wasting resources on low-ranked ones,
similar 
to classic top-$k$ queries~\cite{ilyas08survey}.
However, in contrast to top-$k$, where ``pruning'' techniques based on the \emph{given}
number of returned answers $k$ can be leveraged,\footnote{
Besides the requirement of $k$ being fixed in advance,
older work on top-$k$ joins assumes a cost model that accounts for data access, but not for intermediate results~\cite[Part 1]{tziavelis20tutorial}.}
a ranked-enumeration algorithm does not know the value $k$ in advance.
Instead of pruning,
it can at best \emph{postpone} work on lower-ranked answers,
providing guarantees
\emph{no matter how many answers are eventually returned}.
We are thus interested in the
\emph{Time-To-$k$}, or $\TT(k)$, \emph{for any possible value of $k$}.
This gave rise to the ``any-$k$'' label, quasi an ``anytime top-$k$''
algorithm~\cite{boddy1991anytime, yang2018any, YangRLG18:anyKexploreDB}.

Note that a stricter and popular~\cite{bagan07constenum,idris20dynamic_theta,olteanu15dtrees,DBLP:journals/sigmod/Segoufin15} measure of performance involves combining \emph{preprocessing time} (i.e., $\TT(1)$) with the \emph{worst-case delay between answers}
(i.e., the maximum inter-arrival time).
However, lowering the worst-case delay may have no practical benefit if it does not also improve $\TT(k)$~\cite{capelli23delay,carmeli21ucqs,tziavelis23ranked}.
Adopting $\TT(k)$ allows for situations where a spike in delay is offset 
by shorter delays in \emph{previous} iterations.
An established example where this difference occurs is incremental QuickSort~\cite{paredes06iqs}
which guarantees $\TT(k) = \O(n + k \log k)$, 
but has a linear worst-case delay between answers.

\begin{figure}
\small
\begin{verbatim}
  SELECT Cit1.PaperID, Cit2.PaperID, Cit3.PaperID,
         Cit3.CitedPaperID, Cit1.InflWeight + 
         Cit2.InflWeight + Cit3.InflWeight AS Weight
  FROM Cit Cit1, Cit Cit2, Cit Cit3
  WHERE Cit1.CitedPaperID = Cit2.PaperID AND 
        Cit2.CitedPaperID = Cit3.PaperID
  ORDER BY Weight
\end{verbatim}
\caption{SQL query for ranking chains of highly influential citations.}
\label{fig:sql}
\end{figure}

\begin{figure}
    \centering
    \includegraphics[width=0.7\linewidth]{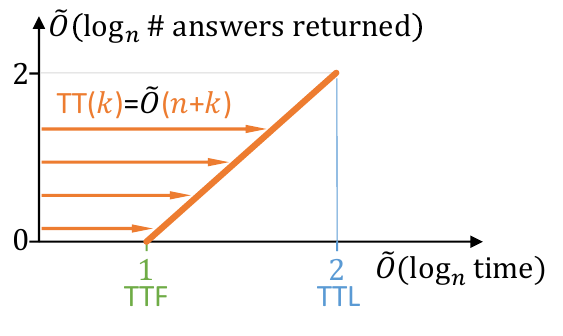}
    \caption{Ranked enumeration guarantees for the query of \Cref{fig:sql}: The first answer (TTF for Time-To-First) is returned in $\Otilde(n)$ and the last answer (TTL for Time-To-Last) in $\Otilde(n^2)$.}
    \label{Fig_enumeration}
\end{figure}

\introparagraph{An Example}
Consider a bibliography dataset that stores
the influence of research papers
on later papers that cite them.
Each tuple in relation 
$\rel{Cit}(\att{PaperID},\allowbreak \att{CitedPaperID},\allowbreak \att{InflWeight})$
states that a paper with ID $\att{CitedPaperID}$ 
influenced a later paper $\att{PaperID}$ 
with a numerical weight $\att{InflWeight}$.
For the sake of the example, assume that the influence weight has been precomputed by some prediction technique
and takes on an integer value in range [1, 10], with 1 being the most influential.
To extract chains of highly influential citations,
we can write the join query
\[
    \rel{Cit}(p_1,p_2,s_1)\,,\,\rel{Cit}(p_2,p_3,s_2)\,,\,\rel{Cit}(p_3,p_4,s_3)
\]
and order its answers in ascending sequence of the SUM $s_1+s_2+s_3$.
For readers unfamiliar with Datalog, 
note that relation $\rel{Cit}$ appears
three times to indicate a self-join
(which requires renaming to $\rel{Cit1},\rel{Cit2},\rel{Cit3}$ in SQL as shown in \Cref{fig:sql}) and that
a variable like $p_2$ appearing more than once indicates an equi-join
between the corresponding columns (i.e., $\rel{Cit1}.p_2 = \rel{Cit2}.p_2$).
How fast can ranked enumeration be here?
The entire query output can have size $n^2$ 
in the worst case~\cite{AGM}.
On the other hand, simply checking if any query answer exists
(called the \emph{Boolean} query)
takes $\Theta(n)$~\cite{DBLP:conf/vldb/Yannakakis81}.
Ranked enumeration aims to cover the continuum between those two
with $\TT(k) = \Otilde(n+k)$, as shown in \Cref{Fig_enumeration}.
The $\Otilde$ notation abstracts away logarithmic factors
in $n$ and $k$
introduced by join indexes
or sorting (by $s_1+s_2+s_3$).

\introparagraph{Prioritizing Computation}
To build intuition, let us first consider how unranked enumeration works. 
If we were to follow a standard table-at-a-time approach, we would start by joining $\rel{Cit1} \bowtie \rel{Cit2}$.
This is a costly bulk computation of time complexity $\Otilde(n^2)$.
However, it would not yet produce a single query answer 
because table $\rel{Cit3}$ has not been checked.
To produce answers as quickly as possible, we need to be more careful in where we spend resources and prioritize differently.
Instead of a table-at-a-time, a \emph{tuple-at-a-time} approach 
is needed. We start with only one tuple from $\rel{Cit1}$, look up the matches in
$\rel{Cit2}$, pick one, and then look up the $\rel{Cit3}$ matches to produce one answer.
This strategy can be implemented using a pipelined execution in a database system.  
The standard unranked enumeration algorithm~\cite{bagan07constenum} achieves $\Otilde(n+k)$
by following such an approach, preceded by a $\Otilde(n)$-time
semi-join reduction~\cite{DBLP:conf/vldb/Yannakakis81},
which removes ``dangling'' tuples that do not contribute to the final output.

Ranked enumeration appears more challenging because additional prioritization is required to avoid low-ranking query answers.
Interestingly, a more careful look at the unranked enumeration algorithm~\cite{bagan07constenum}
reveals that, with appropriate sorting of the input relations,
the output naturally follows a \emph{lexicographic order}.
A lexicographic order is
defined by a sequence of variables, such as $s_1 \LexArr s_2 \LexArr s_3$.
It means that the answers 
are first ordered by variable $s_1$, then by $s_2$, and then by $s_3$
(ORDER BY Cit1.InflWeight, Cit2.InflWeight, Cit3.InflWeight in SQL).
This heavily prioritizes the weight of the first citation in the chain;
a chain with weights $1 \LexArr 10 \LexArr 10$ would be ranked higher than a chain with weights $2 \LexArr 1 \LexArr 1$.
The enumeration algorithm of Bagan et al.~\cite{bagan07constenum} is capable of producing such an order,
granted that we first sort each copy of $\rel{Citi}$ by $\att{InflWeight}$.

But what if a different order that is ``inconvenient'' for the algorithm is required?
As we will discuss in more detail, certain lexicographic orders, such as $s_2 \LexArr s_1 \LexArr s_3$ cannot be achieved by this approach.
Moreover, for SUM ranking, the situation is more difficult because
a high-ranking tuple in $\rel{Cit1}$ might only join with low-ranking tuples in $\rel{Cit2}$ and $\rel{Cit3}$, leading to low-ranking answers in aggregate.
Addressing this requires a stronger form of prioritization that incorporates \emph{lookahead} information about tuples and weights that come later in the query plan.

\introparagraph{Any-$k$ Algorithms}
\revchange{Recent developments led to the design and implementation of any-$k$ algorithms
achieving $\TT(k) = \Otilde(n+k)$
for acyclic join queries and 
appropriately monotone ranking functions~\cite{deep21,tziavelis20vldb}.}
These include \emph{all} lexicographic orders, SUM, as well as MIN and MAX.
In our example, the first $k=\O(1)$ answers are
obtained after only $\Otilde(n)$,
and---if the enumeration is carried out to the end---the last answer in $\Otilde(n^2)$,
matching the join-then-rank approach.
Compared to unranked enumeration, ranking by $s_1+s_2+s_3$ introduces only a
logarithmic factor in $k$.

Although multiple any-$k$ algorithms exist,
their complexity differences
concern logarithmic factors and treating query size as a
variable that can grow arbitrarily,
which may not always materialize in practice.
In this article, we cater to practicality and ease of understanding, focusing on data complexity, guarantees in $\Otilde$ without logarithmic factors,
and on the easiest-to-understand variant.\footnote{
The specific variant we present is \ANYKPART with eager sorting~\cite[][Figure 6]{tziavelis20vldb}.}
We describe the algorithm in a streamlined 
way, without 
the graph abstraction that
\revchange{has been}
used~\cite{tziavelis20vldb}
to highlight the connection to earlier work
on shortest-path enumeration~\cite{jimenez99shortest,lawler72}.

\introparagraph{Organization}
The rest of this article is organized as follows.
\Cref{sec:basic} introduces necessary concepts and notation.
\Cref{sec:lex} presents a simple algorithm that works for certain lexicographic orders
and explores which lexicographic orders are achievable with this algorithm.
\Cref{sec:sum} takes on the harder
case of SUM.
\Cref{sec:extensions} discusses several extensions that generalize the approach
to more expressive queries and ranking functions.
\Cref{sec:conclusion} concludes and provides directions for future work.

\section{Basic Concepts}
\label{sec:basic}

We focus on Select-Project-Join queries, which we formally define in the usual way as Conjunctive Queries.
Throughout the article, we use
$[m]$ to denote the set of integers $\{1, \ldots, m\}$.

\introparagraph{Database}
A \emph{database} $D$ is a set of finite relations $\{ R_1, \ldots, R_m \}$,
where each $R_i$ for $i \in [m]$ 
has arity $\ar(R)$
(i.e., $\ar(R)$ attributes or columns) 
and draws values from a fixed infinite domain $\dom$, 
(i.e., $R_i \subseteq \dom^{\ar(R_i)}$).
The size of the database $n$ is the number of tuples across all relations.

\introparagraph{Query}
In Datalog,
a \emph{Conjunctive Query} (CQ) $Q$ is an expression
$Q(\vec Y)  \allowbreak\datarule\allowbreak R_1(\vec{V_1}),\allowbreak \ldots,\allowbreak R_\ell(\vec{V_\ell})$,
where each $\vec{V}_i$ for $i \in [\ell]$
is a list of either \emph{variables} (representing database attributes)
or constants from $\dom$ (encoding selection).
Each \emph{atom} $R_i(\vec{V_i})$ 
refers to a (not necessarily distinct)
database relation with $|\vec{V_i}|$ attributes.
If $\vec X$ is the set of all distinct variables appearing in all lists $\vec{V}_i$ for $i \in [\ell]$,
then the variables $\vec Y$ (representing output attributes) need to be a subset of $\vec X$ and are called \emph{free}.
A \emph{Join Query} (JQ) is a special case of a CQ where all variables are free
(i.e., $\vec Y = \vec X$).
Multiple atoms are allowed to refer to the same relation,
resulting in a \emph{self-join}.
The query size, measured by the number of symbols in the query,
is assumed to be $\O(1)$.
This is often referred to as \emph{data complexity}~\cite{DBLP:conf/stoc/Vardi82}
and it is relevant in practice because while new data may be collected,
the query size does not typically grow unboundedly.

Queries are evaluated over a database $D$ and produce a result $Q(D)$.
A \emph{query answer}
or \emph{output tuple} is an element $q \in Q(D)$.
The occurrence of the same variable in different atoms encodes an \emph{equi-join} condition,
implying equality between the corresponding attributes.
A typical preprocessing step for all algorithms is to (1) 
remove self-joins from the query by copying database tables
and (2) remove selections on individual relations (like $R(x,1)$ or $R(x,x)$)
by filtering.
These operations take $\O(n)$ and can be ignored because the cost is asymptotically the same as reading the database once.
Afterwards, a naive evaluation strategy to compute $Q(D)$
(which helps to understand the query semantics)
is to
($i$) materialize the Cartesian product of the $\ell$ relations,
($ii$) select tuples that satisfy the equi-joins,
and ($iii$) project on the $\vec{Y}$ attributes.

\introparagraph{Acyclicity}
A CQ is (alpha-)\emph{acyclic}~\cite{baron16acyclic} if
it admits a \emph{join tree}.
A join tree is a rooted tree whose nodes are the query atoms and for each variable $x$, all tree nodes
containing $x$
form a connected subtree.\footnote{For an illustration, please see \url{https://www.youtube.com/watch?v=toi7ysuyRkw&t=340} \cite{tziavelis22tutorial}.}
The acyclicity of a CQ can be tested, and a corresponding join tree can be constructed, in linear time in the query size~\cite{tarjan84acyclic}.

\introparagraph{Ranking}
Ranked enumeration assumes a user-specified \emph{ranking function} that orders the query answers $Q(D)$
by mapping them to a domain $W$ equipped
with a total order $\preceq$.
Ties are broken arbitrarily.
Given a query $Q$, a \emph{lexicographic order} $\lexrank$ is a sequence of 
query variables
$x_1 \LexArr x_2 \LexArr \ldots$,
implying that the answers are first compared by the values of $x_1$,
and if tied by the values of $x_2$, and so on.
A \emph{partial} lexicographic order contains a strict subset of the query variables.
Another case is \emph{SUM}, given by an expression
$f_1(x_1) \allowbreak+\allowbreak f_2(x_2) \allowbreak+\allowbreak \ldots$,
where $f_1,\allowbreak f_2,\allowbreak \ldots$ can be arbitrary, $\O(1)$-computable functions mapping $\dom$ to $\R$.
The ranking may alternatively be defined using values on the database tuples instead of the query variables;
the latter can be reduced to the former in linear time as we will see in more detail in \Cref{sec:sum_preprocess}.

\section{Enumeration by Lexicographic Order}
\label{sec:lex}

We begin with the lexicographic orders that can be produced as a by-product
of the standard ``unranked'' enumeration algorithm
through a minor extension (i.e., pre-sorting all input relations according to the lexicographic order).
Although various descriptions of this algorithm exist in the literature using different
abstractions~\cite{bagan07constenum,Berkholz20tutorial,olteanu15dtrees,DBLP:journals/sigmod/Segoufin15},
it is often overlooked that it can easily produce query answers according to
certain lexicographic orders.

We offer a detailed description that ($i$) is easy to implement
and ($ii$) generalizes to SUM (\Cref{sec:sum}) and other orders.
We focus on acyclic JQs and discuss
how this restriction can be lifted in \Cref{sec:extensions}.

The algorithm consists of two phases. First, the preprocessing phase 
builds essential
data structures such as join indexes
and
applies a semijoin reduction~\cite{DBLP:conf/vldb/Yannakakis81}
to remove dangling tuples from the input relations. 
Then, the enumeration phase traverses the relations
using the indexes to connect joining tuples.
The $\Otilde(n + k)$ complexity guarantee for $\TT(k)$ hinges on the semijoin filtering, which eliminates ``dead-ends'' 
by ensuring that
every partial query answer---generated by joining tuples from a subset of relations---can
be extended to a complete query answer.
We will detail both phases in \Cref{sec:lex_preprocess,sec:lex_enumeration},
then examine, which lexicographic orders can be supported by this algorithm
in \Cref{sec:lex_orders}.

\begin{figure}[t]
    \centering
    \includegraphics[height=5.5cm]{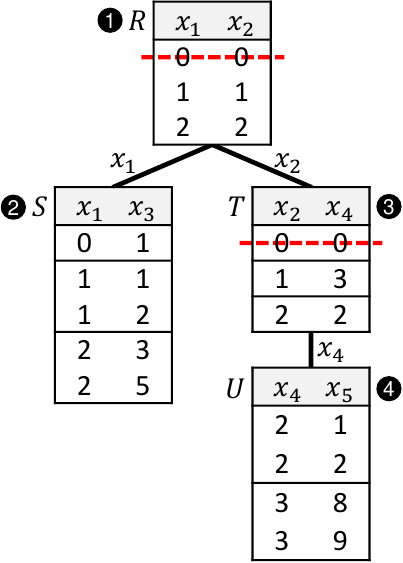}
    \caption{An example database for the join query $R(x_1, x_2), S(x_1, x_3), T(x_2, x_4), U(x_4, x_5)$. The relations are organized in a join tree. Red marks indicate tuples removed by the semijoin reduction.
    Also shown are shared variables between child-parent pairs and the relation ordering $\relind$
    used by the lexicographic enumeration algorithm.
    }
    \label{fig:example}
\end{figure}

As a guiding example, we use the query 
$R(x_1, x_2),\allowbreak S(x_1, x_3),\allowbreak T(x_2, x_4),\allowbreak U(x_4, x_5)$ and show how to achieve the order $x_1 \LexArr x_2 \LexArr x_3 \LexArr x_4 \LexArr x_5$. 
An example database is shown in \Cref{fig:example}.

\subsection{Bottom-up Preprocessing Phase}
\label{sec:lex_preprocess}

\introparagraph{Join Order}
The preprocessing phase starts by organizing the relations in a 
(rooted)
join tree $\calT$.
Unlike a database query plan, a join tree does not fully specify the join order.
It only determines that a parent relation must be processed before its children
(also called a topological sort). Hence any order that respects this constraint
can be followed by the enumeration algorithm.
Let function $\relind$ denote such an order, i.e., it maps the integers $[\ell]$ 
to database relations, where $\ell$ is the number of relations.
In our example, we have $\relind(1) \!=\! R, \relind(2) \!=\! S, \relind(3) \!=\! T, \relind(4) \!=\! U$
(see \cref{fig:example}).
This means that relation $S$ will always be visited before relation $T$ during enumeration.
To encode the tree structure, we refer to the parent of the $r$-th relation 
in the order as $\parent(r)$, for
$r \in [2, \ell]$.

\introparagraph{Join Indexes}
Next, we build join indexes, e.g., B-trees or hash indexes, 
allowing us to
find matching tuples efficiently. 
We abstract an index as a function $\joinindex{R}{S}$,
which, given a tuple $t \in R$,
returns a list $M$ of $S$ tuples that agree with $t$ on the join attributes
between $R$ and $S$ (i.e., the common variables between the atoms).
To achieve the desired $\TT(k)$ guarantees, the index must be built in
$\Otilde(n)$ with lookups in $\Otilde(1)$ (not including the time it takes to read $M$).
We construct one index for each parent-child pair in the join tree,
i.e., $\joinindex{R}{S}$ based on $x_1$, $\joinindex{R}{T}$ based on $x_2$, $\joinindex{T}{U}$ based on $x_4$ in our example.
\Cref{fig:example} shows each relation grouped by the attributes that join with the parent, i.e., the image of $\joinindex{\parent(r)}{\relind(r)}$ for the $r$-th relation,
$r \in [2, \ell]$.
The root has no grouping.

\introparagraph{Semijoin reduction}
Using the join tree and indexes, we perform a \emph{semijoin reduction}
exactly as in the bottom-up step of the Yannakakis algorithm~\cite{DBLP:conf/vldb/Yannakakis81}.
The relations are traversed in reverse topological order with a semijoin
applied for each parent-child pair.
In our example, the semijoins are executed in the following order:
\[
T = T \SemiJoin U,\;\; R = R \SemiJoin T,\;\; R = R \SemiJoin S
\]
This step is crucial for our desired complexity guarantee.
To understand why, consider tuple $R(0, 0)$, for which
there are matching tuples in $S$ and $T$, but none in $U$.
Consequently, the time processing $R(0, 0)$ is wasted, without producing an output
tuple. With sufficiently many such ``dangling'' tuples, the time between consecutive
answers would grow to exceed $\TT(k) = \Otilde(n + k)$.
The semijoin reduction prevents this by removing dangling tuples like
$R(0, 0)$ and $T(0,0)$. 
Notice that $S(0, 1)$ is dangling, but not removed.
Removing all dangling tuples would require a \emph{full reduction}~\cite{BeeriFMY83,DBLP:journals/jacm/BernsteinC81},
which is not necessary for the enumeration algorithm. It is easy to show that
any remaining dangling tuples will never be accessed by top-down traversals.

\Cref{alg:semijoin} presents the semijoin reduction expressed in a way that
easily generalizes to support other orders, as we will see in \Cref{sec:sum}.
Specifically, it can be viewed as message passing at the tuple level:
Each tuple pulls ``messages'' from joining tuples in the children relations,
determines its own state based on the messages, and later passes a message
up the tree. The ``message'' here is a Boolean value that indicates whether
matching tuples exist in the subtree. If the aggregated message from at
least one of the children relations is ``False'', 
then the tuple is removed and a ``False'' message is propagated upwards.
Note that parents are ``pulling'' instead of children ``pushing'' messages so that we can
use the parent-to-child join indexes that we anyway need in the enumeration phase.
The algorithm employs \emph{memoization} for the aggregated message of a join group (\Cref{alg:sj-memoization}),
since multiple tuples in the parent relation may access it.
This is important in order to guarantee linear time.

\introparagraph{Sorting}
When we build a join index, we sort its entries
(i.e., the tuples within the same join group)
by the same lexicographic order. 
In the example,
the entries of $\joinindex{R}{S}$ are sorted
by $x_1 \LexArr x_3$,
the entries of $\joinindex{R}{T}$
by $x_2 \LexArr x_4$,
and so on.
The join index is built after reducing a relation with
messages from its children,
and sorted thereafter.
The tuples of the root relation are considered to belong to the same join group
(as if a parent relation with an empty set of join variables to group-by existed)
and are also sorted.
Slightly abusing the notation,
we treat a relation as a sorted list of tuples;
e.g., $R[1]$ denotes the first tuple of $R$.

\begin{algorithm}[tbp]
\linespread{0.85}\selectfont
\SetAlgoLined
\LinesNumbered
\SetInd{0.5em}{0.8em}
\SetArgSty{upshape}

\nonl
\textbf{Input}: acyclic JQ $Q$ (without self-joins), database $D$, join tree $\calT$, lexicographic order $\lexrank$,
relation ordering $\relind$ consistent with $\calT$\\
\textbf{Output}: reduced and sorted database $D'$, $\joinindex{R}{S}$ for each parent $R$ and child $S$ in $\calT$\\
\vspace{1mm}

Initialize $\tupval(t) = \True$ for all tuples $t$ of all relations\;

\algocomment{Process relations in reverse $\relind$ order (bottom-up
in $\calT$)}

\For{$i = \ell$ down to 2 \label{alg:sj-bottom-up}}
{   
    relation $S = \relind[i]$; relation $R = \parent(S)$

    \algocomment{Relation $S$ has been reduced in a previous iteration\\(or is a leaf)}\;

    Construct $\joinindex{R}{S}$ on shared attributes\;

    Sort $\joinindex{R}{S}$ entries by $\lexrank$\;

    \For{tuple $t \in R$}
    {
            $M = \joinindex{R}{S}(t)$\;
            \algocomment{Memoization: $\tupval(M)$ is reused}\;
            \If{$\tupval(M)$ not already computed\label{alg:sj-memoization}}{
            $\tupval(M) = \False \lor \bigvee_{t' \in M} \tupval(t')$ \label{alg:sj-disj}\;
            }
            $\tupval(t) = \tupval(t) \wedge \tupval(M)$ \label{alg:sj-conj}\;
            \lIf{not $\tupval(t)$}{remove $t$ from $R$ in $D$}
    }
    
}

Sort the root $R$ by $\lexrank$\;

\Return $D$, $\joinindex{R}{S}$ for all $(R, S) \in \calT$\;

\caption{Preprocessing for lexicographic enumeration (\Cref{sec:lex_preprocess}).}
\label{alg:semijoin}
\end{algorithm}

\subsection{Top-down Enumeration Phase}
\label{sec:lex_enumeration}

While the semijoin reduction proceeds bottom-up in the opposite direction of
relation order $\relind$ (\Cref{alg:sj-bottom-up}),
the enumeration phase traverses the relations top-down.
We start with tuple $R[1] = R(1, 1)$ and, through the join indexes, find the first
match in every relation, yielding the first query answer $(1, 1, 1, 3, 8)$.\footnote{Answers
are represented as a tuple of values assigned to variables $(x_1,\ldots, x_5)$,
or alternatively,
as a list of joining tuples $[t_1, \ldots, t_4]$. 
For ease of presentation, we use the former in text and the latter in pseudocode.}
In the second iteration, we proceed with the matches from the last relation, i.e.,
tuple $(3,9)$ from $U$, obtaining $(1, 1, 1, 3, 9)$.
This exhausts all matches in $U$, therefore in the third iteration
the algorithm backtracks to the next match in preceding relation $T$.
Since no second match exists in $T$, we backtrack once again to $S$,
encountering $(1,2)$ there. With $(1,1,2)$ as a partial answer, the algorithm
proceeds forward to $T, U$ to obtain $(1,1,2,3,8)$.
The process continues analogously, returning answers
$(1, 1, 2, 3, 9)$, $(2, 2, 3, 2, 1)$, etc.\footnote{For an illustration, please see \url{https://www.youtube.com/watch?v=toi7ysuyRkw&t=1720s} \cite{tziavelis22tutorial}.}

\begin{figure*}[t]
    \centering
    \includegraphics[width=\linewidth]{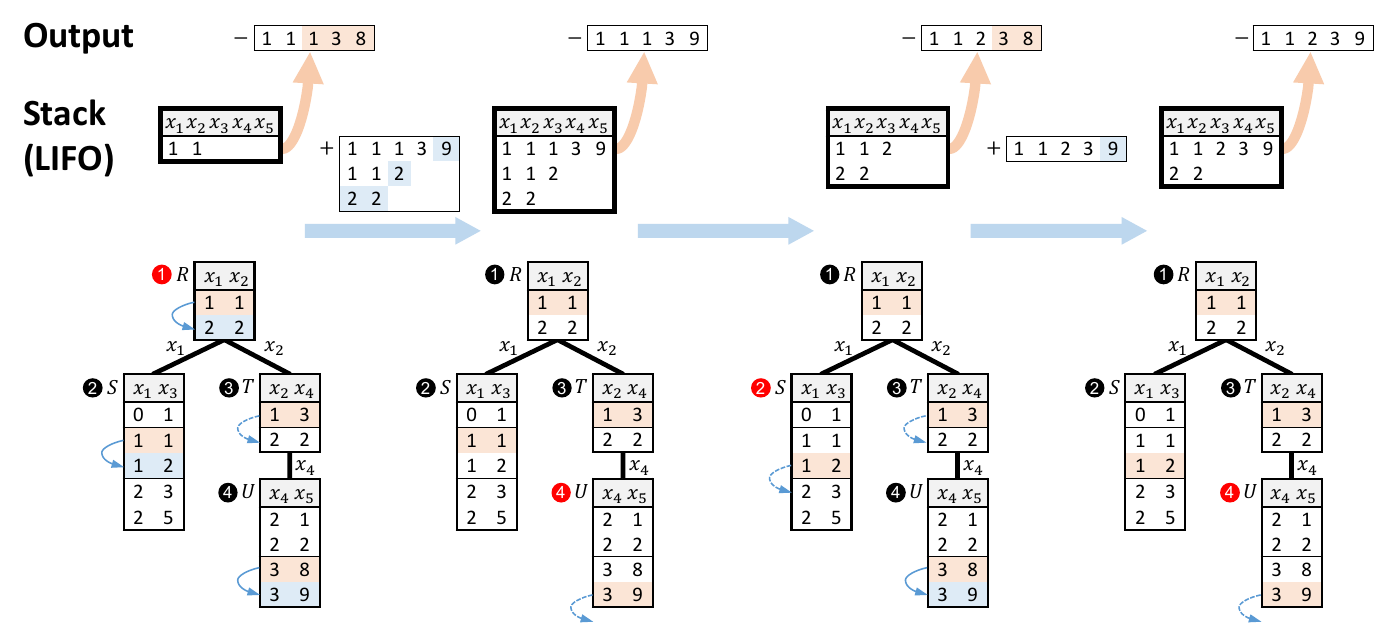}
    \caption{Enumeration steps for the first 4 answers by \Cref{alg:lex} described in \Cref{sec:lex}. 
    The stack, shown on top, pops a partial answer, which is extended with the first matching tuples (in orange color) and moved to the output in each iteration.
    Starting from the last relation for which a tuple is in the partial answer (in red color),
    we check if a ``next'' tuple in the same join group exists (in blue color) and push a new answer to the stack.
    Dashed arrows indicate that there is no next.}
    \label{fig:enumeration}
\end{figure*}

This enumeration can be implemented recursively, akin to a standard depth-first search (DFS).
Equivalently, we implement it with a \emph{stack} 
of partial query answers
(LIFO), 
which tracks the current frontier.
A partial answer contains matched tuples from only a subset of the relations.
When a partial answer is popped from the stack
and we extend it into a complete answer,
alternatives that use the next available tuple are pushed back onto the stack, starting from the current relation.
The extension of a partial answer always selects the first matching tuple following the relation order.
This is illustrated in \Cref{fig:enumeration}.
Notice that when
the second answer
$(1,1,1,3,9)$ is popped, the current relation is $\relind(4) = U$, 
with $R(1,1)$, $S(1,1)$, and $T(1,3)$ from the earlier relations considered fixed
(so we do not generate new answers from those relations).
In fact, $(1,1,2)$ is already on the stack from the previous iteration.
This logic ensures that we enumerate each query answer exactly once.
For the time complexity, note that we visit each relation at most once in each iteration, thus the cost per iteration is $\Otilde(1)$ because
query size is treated as a constant.

The LIFO nature of the stack is essential for achieving the lexicographic order.
For instance, new answers that replace $U$-tuples (thus, change only the $x_5$ value)
are always popped before answers that replace tuples in $R$, $S$, or $T$.
In the following, we discuss the achievable orders in more detail.

\begin{algorithm}[tbp]
\linespread{0.85}\selectfont
\SetAlgoLined
\LinesNumbered
\SetInd{0.5em}{0.8em}
\SetKwFunction{RecFun}{next}
\SetKwProg{Fn}{Function}{:}{}
\SetArgSty{upshape}

\nonl
\textbf{Input}: acyclic JQ $Q$, database $D$, lexicographic order $\lexrank$ without disruptive trio\\
\nonl
\textbf{Output}: Ranked enumeration of $Q(D)$ in $\lexrank$ order \\
\vspace{1mm}

Remove self-joins from $Q$ by copying the corresponding relations and renaming them in both $D$ and $Q$\;

Construct an $\lexrank$-consistent join tree $\calT$ of $Q$
with $\lexrank$-consistent relation order given by 
$\relind(i), i \in [\ell]$\;

Preprocess($Q$, $D$, $\calT$, $\lexrank$, $\relind$) (\Cref{alg:semijoin})\;

\algocomment{A partial answer in the stack is represented as a list of input tuples together with their positions in the corresponding join groups (to easily get the next)}\;
Initialize stack $\calS$ with element $[ (t_1, 1) ]$ 
where $t_1 = R[1]$ and $R$ is the root of $\calT$\;

\Repeat {query is interrupted or $\calS$ is empty}{\label{line:lex_repeat}

    \algocomment{Pop a partial answer ($1 \leq r \leq |\calT|$),
    which also contains the positions $j_i$ for each tuple $t_i$}\;
    $s = \calS.\mathrm{pop}()$; $[ (t_1, j_1), \ldots, (t_r, j_r) ] = s$\;

    \algocomment{Look up matches in $r$-th relation}\;
    $M_r = \joinindex{\parent(r)}{\relind(r)}(t_{\parent(r)})$\;

    \algocomment{Push partial answer with next tuple of $r$-th relation. It exists if $j_r$ is not the last position in the group $M_r$}\;
    \If{$|M_r| \geq j_{r}+1$}
    {
        $s' = s.\mathrm{copy}().\mathrm{replaceLast}((M_r[j_{r}+1], j_{r}+1))$ \label{line:lex_replace1}\;
        $\calS.\mathrm{push}(s')$\;
    }

    \algocomment{Range over the remaining relations}\;
    \For{$i$ from $r+1$ to $|\calT|$}{

        \algocomment{Look up matches in $i$-th relation}\;
        $M_i = \joinindex{\parent(i)}{\relind(i)}(t_{\parent(i)})$\;

        \algocomment{Extend partial answer with first tuple in matches of $i$-th relation}\;
        $t_i = M_i[1]$; $s.\mathrm{append}((t_i, 1))$\;
        
        \algocomment{$s$ is now $[ (t_1, j_1), \ldots, (t_r, j_r), \ldots, (t_i, 1)]$}\;

        \If{$|M_i| \geq 2$}
        {
            \algocomment{Push partial answer with next tuple of $i$-th relation to stack $\calS$}\;
            $s' = s.\mathrm{copy}().\mathrm{replaceLast}((M_i[2], 2))$ \label{line:lex_replace2}\;
            $\calS.\mathrm{push}(s')$\;
        }

    }
    
    Merge $s$ into single tuple and output\label{line:lex_output}\;
}

\caption{Ranked enumeration for lexicographic orders without disruptive trios}
\label{alg:lex}
\end{algorithm}

\subsection{Supported Lexicographic Orders}
\label{sec:lex_orders}

Different lexicographic orders can be achieved by different sortings of the individual relations.
For example, if we sort $R$ by $x_2 \LexArr x_1$, we can achieve the order $x_2 \LexArr x_1 \LexArr x_3 \LexArr x_4 \LexArr x_5$
without any other change in the algorithm.
Some other orders can be achieved by additionally selecting a different topological sort on the join tree.
With $[R,T,U,S]$ instead of $[R,S,T,U]$, we can achieve $x_1 \LexArr x_2 \LexArr x_4 \LexArr x_5 \LexArr x_3$.
However, certain lexicographic orders cannot be achieved by this algorithm.
Brault-Baron~\cite{brault13thesis} identified a sufficient condition,
which was later termed a \emph{disruptive trio}~\cite{carmeli23direct} and 
shown to be necessary for other problems related to enumeration. (We discuss this in more detail in \Cref{sec:da}.)

\begin{definition}[Disruptive Trio]
\label{def:trio}
For a CQ $Q$ and lexicographic order $\lexrank$, three variables $x_1, x_2, x_3$ from $\lexrank$ 
with relative order $x_1 \LexArr x_2 \LexArr x_3$
form a disruptive trio 
if $x_1$ and $x_2$ are not neighbors (i.e., they do not appear together in a $Q$ atom), 
but $x_3$ is a neighbor of both $x_1$ and $x_2$.

\end{definition}

In our example, $x_1, x_4, x_2$ form a disruptive trio if $\lexrank$ is
$x_1 \LexArr x_4 \LexArr x_2$ or even $x_1 \LexArr x_3 \LexArr x_4 \LexArr x_5 \LexArr x_2$.
Intuitively, during the enumeration, we cannot transition from $R$ to $T$
without fixing $x_1, x_2$ before $x_4$, which is inconsistent with the order $\lexrank$.

Brault-Baron showed that for lexicographic orders containing all free variables,
the absence of a disruptive trio
is equivalent to $\lexrank$ being a \emph{reverse alpha elimination order}~\cite[][Theorem 15]{brault13thesis},
and for partial lexicographic orders, it is equivalent to the lexicographic order being \emph{consistent} with (or, in other words, a restriction of) a reverse alpha elimination order.
An alpha elimination order is an ordering of the variables that guides the join tree construction~\cite{baron16acyclic}.
If variable $y$ follows variable $x$ in the elimination order,
then in the resulting join tree,
$y$ will never appear without $x$ in any ancestor of a node that contains $x$.\footnote{A similar property has been proposed in factorized databases in order to detect whether a lexicographic order is admissible with a given factorization order~\cite{bakibayev13fordering}.}
This guarantees that there exists a relation ordering $\relind$
such that the $\lexrank$ variables are encountered in the desired sequence.
We call such an ordering of the relations, and its corresponding join tree, \emph{$\lexrank$-consistent}.

If a desired lexicographic order has no disruptive trio, then we can find an $\lexrank$-consistent join tree and an $\lexrank$-consistent ordering of the relations
to use with the enumeration algorithm discussed above.

\begin{theorem}[LEX]
\label{th:lex}
Let $Q$ be an acyclic join query over database $D$ and $\lexrank$ a lexicographic order of the variables in $Q$.
If $\lexrank$ does not contain a disruptive trio, then
ranked enumeration of $Q(D)$ by $\lexrank$ can be achieved with $\TT(k) = \Otilde(n + k)$.
\end{theorem}

\Cref{alg:lex} shows the pseudocode. After the preprocessing phase,
a loop returns query answers iteratively by popping and pushing from the stack.
Notice that, for each answer, the algorithm keeps track of the positions $j_1, \ldots, j_\ell$ of the tuples
within the corresponding join group.
This allows it to quickly access the next tuple in the group when constructing new answers (in \Cref{line:lex_replace1,line:lex_replace2}).

What about the lexicographic orders that contain disruptive trios?
\Cref{alg:lex} does not apply because there is no join tree that can match the order.
For these orders, as well as SUM, we need a different strategy.
Lexicographic orders with disruptive trios can in fact be
reduced to a SUM-ordering problem by assigning the appropriate variable weights:
If all relations have cardinality at most $n$,
we can achieve that by setting the weight of the $i^\textrm{th}$ value of the $j^\textrm{th}$
variable in the order to $i \cdot n^{|\lexrank|-1-j}$.

\section{Enumeration by SUM Order}
\label{sec:sum}

In this section, we shift focus to ranking by SUM.
Let $\sum_{i=1}^5 x_i$ (in ascending order) be the ranking function for our example query.
A naive strategy is to select the best tuple from each relation based on its individual weight.
For instance, using the same join tree as before, 
we could start with $R(1,1)$ since it has the lowest weight $1+1=2$ within $R$.
However, this strategy is not guaranteed to find the top answers,
at least not within the time bounds we aim for. Once we choose $R(1,1)$, 
we will be stuck in a region of query answers with high overall weight because of the
high weights of $U(3, 8)$ and $U(3, 9)$,
which are the only matching tuples in $U$.
The true top-1 answer $(2, 2, 3, 2, 1)$ starts with $R(2, 2)$, which matches with $U(2, 1)$.
To make the right choices in $R$, the algorithm needs ``lookahead'' information
about later matches in relations like $U$.

Unfortunately, it is infeasible to explicitly pre-compute the ``lookahead''
combinations of $S, T, U$ tuples in the preprocessing phase, 
because that would exceed our desired $\Otilde(n)$.
Instead, we rely on Dynamic Programming and a factorized representation
of the query output. The enumeration phase is similar to the algorithm of
\Cref{sec:lex_enumeration}, but uses a \emph{priority queue} instead of a \emph{stack} in order
to prioritize the candidates according to the ``lookahead'' information computed
during preprocessing.

\subsection{Bottom-up Preprocessing Phase}
\label{sec:sum_preprocess}

To prioritize the tuples that lead to the lowest total weight,
we modify the semijoin reduction so that, in addition to removing dangling tuples,
we also compute
the \emph{best possible weight} $\tupopt(t)$ reachable by each tuple $t$ when joining it with other tuples in its subtree.
This bottom-up computation is essentially a form of Dynamic Programming.

\begin{figure}[t]
    \centering
    \includegraphics[height=5cm]{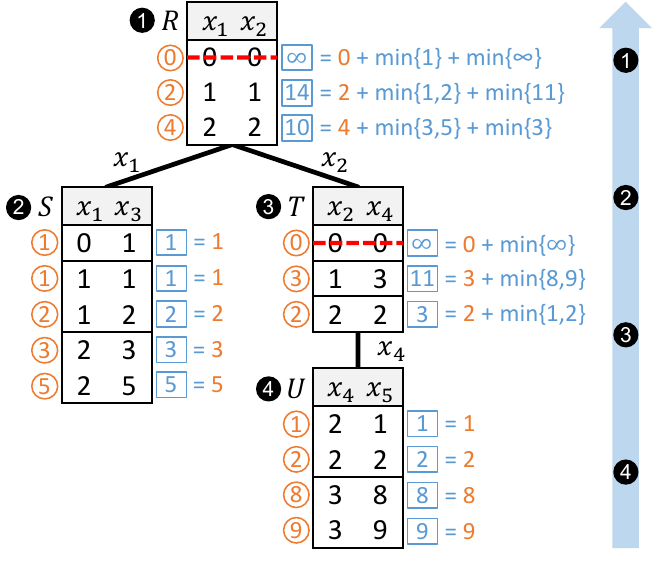}
    \caption{
        Bottom-up Dynamic Programming for SUM (preprocessing).
         Orange circles show the weights $\wt(t)$ assigned to tuples. 
    Blue squares show the calculated minimal subtree weight $\tupopt(t)$ for each tuple.
    }
    \label{fig:dp}
\end{figure}

The algorithm is easier to present using tuple weights instead of attribute weights.
We set the weights of $R$ to $x_1+x_2$, of $S$ to $x_3$, of $T$ to $x_4$ and of $U$ to $x_5$.
Such a conversion is always possible in linear time, which means that both regimes are supported in the algorithm.
We only need to be careful so that the weight of each variable is assigned to a unique relation;
this can be achieved through a mapping $\mu$ that assigns each variable $x$ in the SUM
to the first relation (or atom) that contains $x$ in the topological sort $\relind$.
We denote the weight of tuple $t$ by $\wt(t)$.
\Cref{alg:dp} computes $\tupopt(t)$ for all tuples $t$ by
aggregating the input weights using $\min$ and $+$, bottom-up
in reverse $\relind$ order
$\circled{4} \LexArr \circled{3} \LexArr \circled{2} \LexArr \circled{1}$, as shown in \Cref{fig:dp}.
The leaf relations set $\tupopt(t)$ to be equal to $\wt(t)$.
For tuple $T(2,2)$, which is in the non-leaf relation $T$,
we add its own weight $2$ to the message $\min\{1,2\}$ from the joining group in $U$, 
hence $\tupopt(T(2, 2)) =\! 2 \!+\! \min\{1, 2\} =\! 3$.
For a relation with multiple children, we add the messages from all of them.
E.g., for $R(2, 2)$, we add its own weight $4$ with the message $\min\{3,5\}$ from $S$
and the message $\min\{3\}$ from $T$, hence $\tupopt(R(2, 2)) = 10$.
By the end of the preprocessing step, we know the optimal weight
$\tupopt(t)$ for each tuple $t$,
and the join index entries are \emph{sorted} according to these values.

\begin{remark}
The fact that \Cref{alg:dp} is so similar to the semijoin reduction in \Cref{alg:semijoin} is not a coincidence.
They are both instances of the FAQ framework~\cite{abo16faq}
with different semirings.
In particular, the aggregation operators $\vee$ and $\wedge$ 
from the semi-join reduction
are replaced with $\min$ and $+$
in the variant for SUM.
In more technical terms, the former corresponds to the Boolean semiring and the latter to the tropical semiring.
\end{remark}

\begin{algorithm}[tbp]
\linespread{0.85}\selectfont
\SetAlgoLined
\LinesNumbered
\SetInd{0.5em}{0.8em}
\SetArgSty{upshape}

\nonl
\textbf{Input}: acyclic JQ $Q$ (without self-joins), database $D$ \change{with tuple weights}, join tree $\calT$, relation ordering $\relind$ consistent with $\calT$\\
\textbf{Output}: reduced and sorted database $D'$, $\joinindex{R}{S}$ for each parent $R$ and child $S$ in $\calT$\\
\vspace{1mm}

Initialize \change{$\tupopt(t) = \wt(t)$} for all tuples $t$ of all relations\; 

$\relind =$ relation ordering consistent with $\calT$

\algocomment{Process relations in reverse $\relind$ order (bottom-up
in $\calT$)}

\For{$i = \ell$ down to 2 \label{alg:dp-bottom-up}}
{   
    relation $S = \relind[i]$; relation $R = \parent(S)$\;

    \algocomment{Relation $S$ has been reduced in a previous iteration\\(or is a leaf)}\;

    Construct $\joinindex{R}{S}$ on shared attributes\;

    Sort $\joinindex{R}{S}$ entries by $\tupopt$\;
    
    \For{tuple $t \in R$}
    {
            $M = \joinindex{R}{S}(t)$\;
            \algocomment{Memoization: $\tupval(M)$ is reused}\;
            \If{\change{$\tupopt(M)$} not already computed\label{alg:dp-memoization}}{
            
            \change{$\tupopt(M) = \min\{ \infty, \min_{t' \in M} \tupopt(t')$ \} }\label{alg:dp-min}\;
            }
            \change{$\tupopt(t) = \tupopt(t) + \tupopt(M)$} \label{alg:dp-add}\;
            \lIf{\change{$\tupopt(t) == \infty$}}{remove $t$ from $R$ in $D$}
    }
}

Sort the root $R$ by $\tupopt$\;

\Return $D$, $\joinindex{R}{S}$ for all $(R, S) \in \calT$\;

\caption{Preprocessing for enumeration by SUM\\(\Cref{sec:sum_preprocess}). Changes compared to \Cref{alg:semijoin} are in blue.}
\label{alg:dp}
\end{algorithm}

\begin{algorithm}[tbp]
\linespread{0.85}\selectfont
\SetAlgoLined
\LinesNumbered
\SetInd{0.5em}{0.8em}
\SetKwFunction{RecFun}{next}
\SetKwProg{Fn}{Function}{:}{}
\SetArgSty{upshape}

\nonl
\textbf{Input}: acyclic JQ $Q$, database $D$, SUM order $\sumrank$\\
\nonl
\textbf{Output}: Ranked enumeration of $Q(D)$ in $\sumrank$ order \\
\vspace{1mm}

Remove self-joins from $Q$ by copying the corresponding relations and renaming them in both $D$ and $Q$\;

\change{
Construct a join tree $\calT$ of $Q$ with tree-consistent relation order given by 
$\relind(i), i \in [\ell]$\;

Convert attribute weights to tuple weights\;
}

\change{Preprocess($Q$, $D$, $\calT$, $\relind$) (\Cref{alg:dp})}\;

\algocomment{A partial answer in the stack is represented as a list of input tuples together with their positions in the corresponding join groups (to easily get the next)}\;
Initialize \change{priority queue $\calP$} with element $[ (t_1, 1) ]$ where $t_1 = R[1]$ and $R$ is the root of $\calT$\;

\Repeat {query is interrupted or $\calP$ is empty}{\label{line:sum_repeat}

    \algocomment{Pop a partial answer ($1 \leq r \leq |\calT|$),
    which also contains the positions $j_i$ for each tuple $t_i$}\;
    $s = \calS.\mathrm{pop}()$; $[ (t_1, j_1), \ldots, (t_r, j_r) ] = s$\;

    \algocomment{Look up matches in $r$-th relation}\;
    $M_r = \joinindex{\parent(r)}{\relind(r)}(t_{\parent(r)})$\;

    \algocomment{Push partial answer with next tuple of $r$-th relation. It exists if $j_r$ is not the last position in the group $M_r$}\;

    \If{$|M_r| \geq j_{r}+1$}
    {
        $s' = s.\mathrm{copy}().\mathrm{replaceLast}((M_r[j_{r}+1], j_{r}+1))$ \label{line:sum_replace1}\;
        $\change{\calP}.\mathrm{push}(s')$ \change{with priority $\prio(s')$}\;
    }

    \algocomment{Range over the remaining relations}\;
    \For{$i$ from $r+1$ to $|\calT|$}{

        \algocomment{Look up matches in $i$-th relation}\;
        $M_i = \joinindex{\parent(i)}{\relind(i)}(t_{\parent(i)})$\;

        \algocomment{Extend partial answer with first tuple in matches of $i$-th relation}\;
        $t_i = M_i[1]$;  $s.\mathrm{append}((t_i, 1))$\;
        
        \algocomment{$s$ is now $[ (t_1, j_1), \ldots, (t_r, j_r), \ldots, (t_i, 1)]$}\;

        \If{$|M_i| \geq 2$}
        {
            \algocomment{Push partial answer with next tuple of $i$-th relation to \change{priority queue $\calP$}}\;
            $s' = s.\mathrm{copy}().\mathrm{replaceLast}((M_i[2], 2))$ \label{line:sum_replace2}\;
            $\change{\calP}.\mathrm{push}(s')$ \change{with priority $\prio(s')$}\;
        }

    }
    
    Merge $s$ into single tuple and output\label{line:sum_output}\;
}

\caption{Ranked enumeration for SUM orders. Changes compared to \Cref{alg:lex} are shown in blue.}
\label{alg:sum}
\end{algorithm}

\subsection{Top-down Enumeration Phase}

As can be seen in \Cref{alg:sum}, the high-level logic of the enumeration
is the same as the lexicographic enumeration of \Cref{alg:lex}.
\Cref{line:sum_replace1,line:sum_replace2} generate new query answers with
the tuple in the next position in the join group, like before.
However, tuples within a join group are now sorted by $\tupopt$, 
so
each answer generated is guaranteed to produce the next-best weight
(among those in the same join group) when extended to a complete answer.

\begin{figure*}[t]
    \centering
    \includegraphics[width=\linewidth]{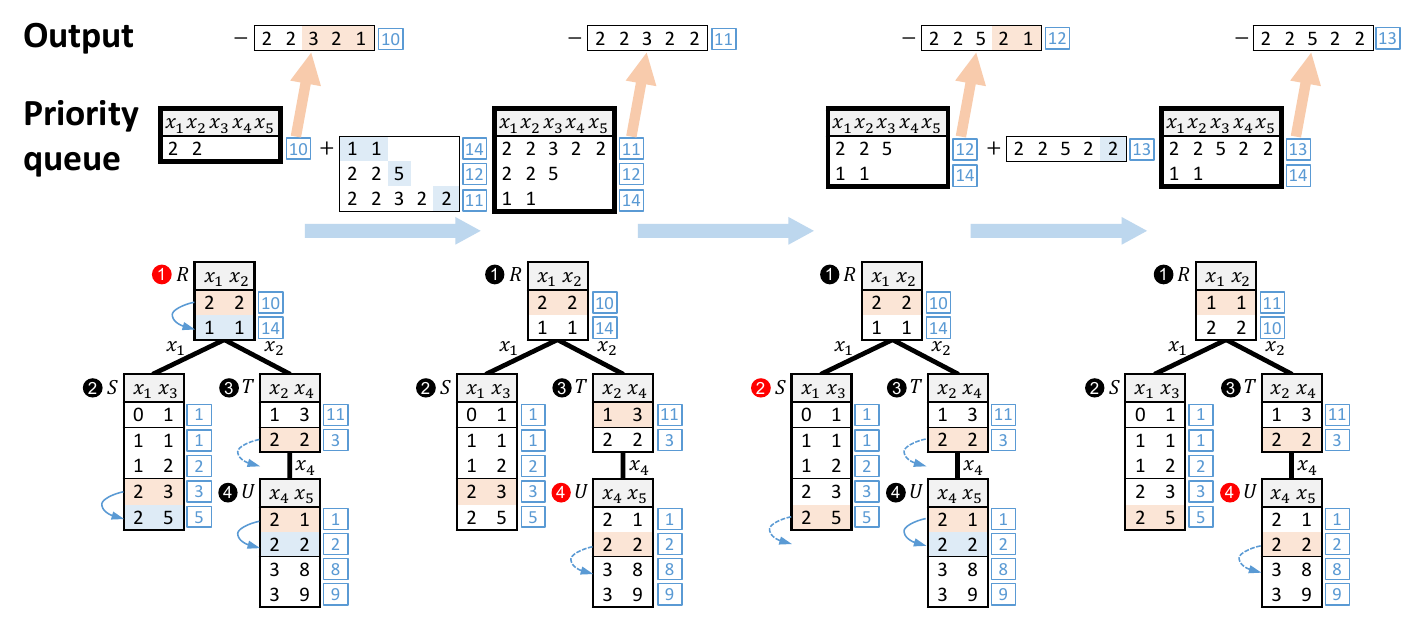}
    \caption{The example from \cref{fig:enumeration} adapted with a priority queue instead of a stack to support ranked enumeration by SUM.}
    \label{fig:sum_enum}
\end{figure*}

Another important difference is that the stack that maintains the partial answers
is replaced by a \emph{priority queue} $\calP$.
Initially, $\calP$ only contains a partial answer with $R[1] = R(2,2)$, producing the top-1 answer $(2, 2, 3, 2, 1)$ with weight $\tupopt(R(2,2)) = 10$.
The second iteration has 3 candidates in $\calP$:
$(1, 1)$, $(2, 2, 5)$ and $(2, 2, 3, 2, 2)$.
The priority of each candidate $s$, denoted by $\prio(s)$ is the weight of the
answer we will obtain if we fully extend it.
We compute it before inserting it into $\calP$;
we can either prematurily extend it into a full answer,
or we can subtract from the previous answer the weight of the subtree that was removed and add the weight of the new subtree.
For example, for $(2, 2, 5)$, we can subtract the weight of the subtree rooted at $S(2,3)$ and add the new weight $\tupopt(S(2,5))$ to the weight of the answer of the previous iteration,
yielding $10-3+5=12$.
Based on the priorities, $(2, 2, 3, 2, 2)$ with priority $11$ will be the winner in the second iteration, and the enumeration continues accordingly.
\Cref{fig:sum_enum} depicts the process.

The size of the priority queue $\calP$ is at most $k\ell$, since we push at most
$\ell$ candidates in each iteration. Hence, the time of each iteration
now includes a logarithmic cost for priority-queue operations (instead of the earlier constant
one for stack accesses). However, if we ignore logarithmic factors,
the $\TT(k)$ complexity remains the same as in \Cref{th:lex}.

\begin{theorem}[SUM]
Let $Q$ be an acyclic join query over database $D$ and $\sumrank$ a SUM ranking function.
Ranked enumeration of $Q(D)$ by $\sumrank$ can be achieved with $\TT(k) = \Otilde(n + k)$.
\end{theorem}

\subsection{\revchange{Performance in Practice}}

\begin{figure}[h]

    \centering
    \begin{subfigure}{\linewidth}
        \centering
        \includegraphics[height=0.6cm]{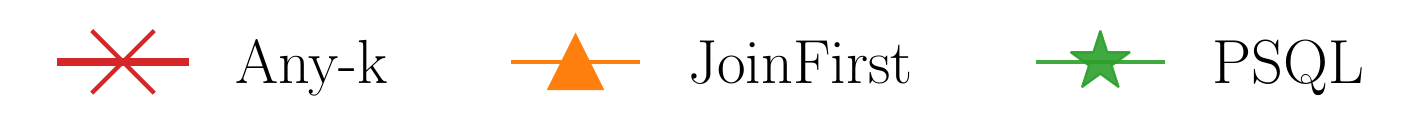}
    \end{subfigure}
    \vspace{-5.5mm}

    \begin{subfigure}{\linewidth}
        \centering
        \includegraphics[width=0.6\linewidth]{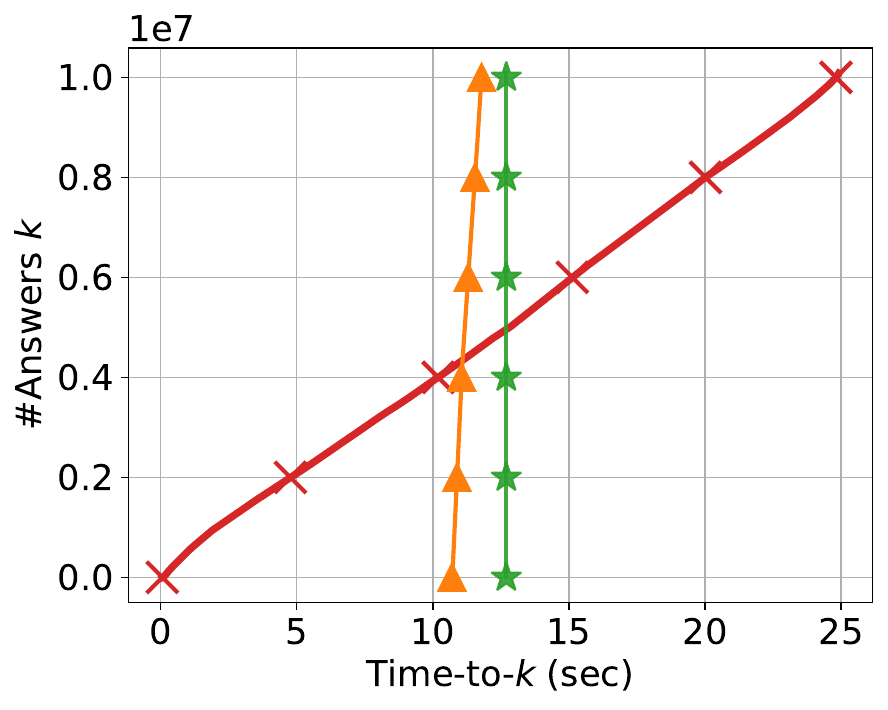}
    \end{subfigure}%

    \caption{Any-$k$ against the join-then-rank approach (\BATCH) and PostgreSQL (\PSQL):
    Any-$k$ returns the top answer in 67 msec, whereas \BATCH needs 10.7 sec~\cite{tziavelis20vldb}.
    }
    \label{exp}

\end{figure}

\revchange{
Any-$k$ (enumeration by SUM) 
has been implemented
and the experimental results from PVLDB'20 \cite{tziavelis20vldb} and PVLDB'21 \cite{tziavelis21inequalities} have been independently reproduced.\footnote{
The code is available to use at \url{https://github.com/northeastern-datalab/anyk-code}.}
In \Cref{exp}, we repeat and show an experiment from PVLDB'20~\cite{tziavelis20vldb}
that measures $\TT(k)$ for a 4-path query (joining relations in a chain) on synthetic data.

The experiment compares
\circled{1} Any-$k$ against
\circled{2} \BATCH (computing the full result with the Yannakakis algorithm~\cite{DBLP:conf/vldb/Yannakakis81}),
and
\circled{3} \PSQL (PostgreSQL 9.5.20).
$\TT(k)$ is depicted on the x-axis and $k$ on the y-axis.
By the time \BATCH returns the first answer (in 10.7 sec),
any-$k$ has already returned more than 4 million, starting with the first one after 67 msec.
\PSQL follows an approach similar to \BATCH and is outperformed for the top-ranked answers.
For the last answer, any-$k$ is slower by less than a factor of 3.
}

\section{\revchange{More General Queries and Tasks}}
\label{sec:extensions}

\revchange{
We review a number of
generalizations that
have been studied,
going beyond the task of ranked enumeration by SUM for acyclic JQs.
}

\subsection{General Ranking Functions}

Beyond lexicographic orders and SUM, the algorithm of \Cref{sec:sum} can be used with any ranking function that obeys a property called \emph{subset-monotonicity}.
Recall that a ranking function $w$ maps the query answers to a domain $W$ 
ordered by $\preceq$.
We consider ranking functions that achieve this by aggregating (a multiset of) input weights via an aggregate function $w_A$.
For example, $w_A$ is $\sum$ for SUM.

\begin{definition}[Subset-Monotonicity]
\label{def:smonotone}
A ranking function $w$ is subset-monotone if
$w_A(X_1) \preceq w_A(X_2) \Rightarrow w_A(X_1 \uplus Y) \preceq w_A(X_2 \uplus Y)$ for all $X_1, X_2, Y$
$\in \N^W$, where $\uplus$ is multiset union.
\end{definition}

Intuitively, subset-monotonicity allows to infer the ranking of complete solutions from the ranking of
partial solutions.
This is essentially enabling Dynamic Programming~\cite{bellman1954}.
Any subset-monotone ranking function can be handled efficiently with the $\Otilde(n+k)$ guarantee for acyclic queries~\cite{tziavelis24thesis,tziavelis23ranked}.\footnote{Alternatively, the ranking function can be defined as a selective dioid~\cite{tziavelis20vldb}, which can be shown to obey subset-monotonicity.}
For instance, we may choose the aggregate function to be $\max$ 
instead of $\sum$.
Under this ranking, only the highest weight is relevant for ordering the answers.

\revchange{
Deep and Koutris~\cite{deep21} generalize subset-monotonicity\footnote{Subset-monotonicity is also referred to as a ``totally decomposable ranking''~\cite{deep21}.} so that the property is sensitive to the join tree structure;
to achieve the desired guarantee,
the property needs to hold
only across the specific nodes of the join tree used by the algorithm.
As an example, consider $f(x,y) + g(z)$ for arbitrary $f, g$ and the query $Q(x,y,z)\allowbreak \datarule \allowbreak  R(x,y), \allowbreak S(y,z)$.
Even though this ranking function is not subset-monotone,
it can be supported efficiently because
$x$ and $y$ are encountered together.
}

What about other ranking functions?
A known negative result is that if the ranking function is a black box, then
one cannot do better than materializing the entire query output~\cite{deep21}.
Thus, the only guarantee we can hope for is the worst-case output size of the query,
given by the AGM bound~\cite{AGM}.

\subsection{CQs with Projection}

So far, we have focused on join queries, yet CQs may also contain projection.
Projections introduce a new challenge: even if ranked enumeration is efficient for a join query,
this may not be true for projections, because we need to eliminate duplicates
(under set semantics), potentially increasing $\TT(k)$. 

Bagan et al.~\cite{bagan07constenum} established a dichotomy for
unranked enumeration that precisely characterizes queries that admit $\TT(k) = \Otilde(n + k)$.
The negative side of the dichotomy applies only to self-join-free CQs and relies on two complexity-theoretic hypotheses:
\sparseBMM{}~\cite{Berkholz20tutorial}
states that two Boolean matrices $A$ and $B$, represented as lists of non-zeros, 
cannot be multiplied in time $m^{1+o(1)}$
where $m$ is the number of non-zeros in $A$, $B$, and $AB$.
\hyperclique{}~\cite{abboud14conjectures,DBLP:conf/soda/LincolnWW18} states that for every $k \geq 2$, there is no
$O(n \polylog n)$ 
algorithm to decide the existence of a
$(k{+}1,k)$-hyperclique in a $k$-uniform hypergraph with $n$ hyperedges,
where a \emph{$(k{+}1,k)$-hyperclique} is a set of $k{+}1$ vertices
such that every subset of $k$ vertices forms a hyperedge,
and a $k$-uniform hypergraph is one where all hyperedges contain exactly $k$ vertices.
Under these assumptions, the only efficient (self-join-free) CQs are those that are \emph{free-connex}.
A CQ is free-connex if it is acyclic and additionally,
it remains acyclic if we add an atom that contains all free variables~\cite{brault13thesis}.

Interestingly, that frontier of tractability for unranked enumeration
turns out to be the same for ranked enumeration with subset-monotone ranking functions
(modulo logarithmic factors).

\begin{theorem}[Dichotomy~\cite{tziavelis24thesis,tziavelis23ranked}]
\label{theorem:cq_data_comp}
Let $Q$ be a CQ.
If $Q$ is free-connex, then ranked enumeration with a subset-monotone ranking function
is possible with $\TT(k) = \Otilde(n + k)$.
Otherwise, if it is also self-join-free, 
then it is not possible with $\TT(k) = \Otilde(n + k)$
for any ranking function,
assuming \sparseBMM{} and \hyperclique{}.
\end{theorem}

For the class of acyclic but non-free-connex CQs, the dichotomy precludes the existence of an algorithm with the efficient $\Otilde(n + k)$ guarantee.
However, $\Otilde(n \cdot k)$ is possible for subset-monotone ranking functions.
This result has been established by the algorithm of Bagan et al.~\cite{bagan07constenum} for lexicographic orders, by Deep et al.~\cite{deep22ranked} for lexicographic orders and SUM through a different algorithm,
and by Kimelfeld and Sagiv~\cite{KimelfeldS2006} for all subset-monotone ranking functions through a third algorithm.

\begin{theorem}[Non-free-connex~\cite{bagan07constenum,deep22ranked,KimelfeldS2006}]
Let $Q$ be an acyclic, non-free-connex CQ.
Ranked enumeration of $Q(D)$ with a subset-monotone ranking function is possible with $\TT(k) = \Otilde(n \cdot k)$.
\end{theorem}

\subsection{Beyond Acyclic CQs}
\label{sec:beyond_acyclic}

We can apply the ranked-enumeration algorithms even to queries that are not acyclic CQs,
albeit with adjusted complexity guarantees. 
This is possible if the query can be transformed into an acyclic and free-connex CQ, or a union of such queries.
In that case, we first apply the transformation and then perform ranked enumeration on the resulting
queries. To deal with a union, we maintain a top-level priority queue that retrieves
the next query answer from the query with the lowest weight in each iteration. 
Duplicate answers introduce potential complications,
but as long as the number of duplicates per answer is bounded by a constant,
they can be filtered on-the-fly without increasing complexity. 
In general, identifying such transformations is an orthogonal research problem, 
and we discuss three notable cases.

\introparagraph{Cyclic JQs}
For cyclic JQs, we can employ (hyper)tree decompositions~\cite{GottlobGLS:2016}
to reduce them to 
a union of
acyclic JQs.
A decomposition is associated with
a width parameter that captures the degree of acyclicity of the query
and affects the complexity of subsequent algorithms; 
for a JQ with width $d$, we can achieve $\TT(k) = \Otilde(n^d + k)$.
The state-of-the-art width for a JQ $Q$ is the submodular width $\subw(Q)$~\cite{khamis17panda,Marx:2013:THP:2555516.2535926},
transforming a cyclic JQ over a database of size $n$ to a union of acyclic JQs of size $\O(n^{\subw(Q)})$,
allowing ranked enumeration with $\TT(k) = \Otilde(n^{\subw(Q)} + k)$.\footnote{An analog exists for CQs (with projection)~\cite{berkholz19submodular}.}

\introparagraph{Built-in Predicates}
Another case involves acyclic JQs that additionally contain built-in predicates~\cite{ullman88book} such as inequalities.
For \emph{non-equalities} (or ``disequalities'' $\neq$),
we can always achieve $\TT(k)=\Otilde(n+k)$ regardless of where the non-equalities appear in the JQ through a ``color-coding'' technique~\cite{papadimitriou99complexity}.
Abo Khamis et al.~\cite{khamis19negation} showed that the same is true for a multidimensional generalization of non-equality, called a Not-All-Equal (NAE) predicate.
For \emph{inequalities} ($<,>$), we can successfully reduce the query to an acyclic JQ 
over an $\Otilde(n)$ database, hence achieving $\TT(k)=\Otilde(n+k)$,
as long as the inequality predicate involves variables that appear in join-tree nodes that are adjacent~\cite{tziavelis21inequalities}.
This condition can be checked directly from the query structure;
it is equivalent to the absence of a chordless path of length at least 4 connecting the inequality variables, in the query's hypergraph~\cite{tziavelis23quantiles}.

\introparagraph{CQs with FDs}
While a CQ may be acyclic but not free-connex, or even cyclic, it may still be possible to transform it to an acyclic CQ without employing a hypertree decomposition, which generally increase the complexity.
This is the case when Functional Dependencies (FDs) are present in the CQ.
We can achieve $\TT(k)=\Otilde(n+k)$ for queries whose so-called \emph{FD-extension},
also known as the \emph{closure} of $Q$~\cite{Gatterbauer2017},
is free-connex~\cite{DBLP:journals/mst/CarmeliK20}.

\revchange{
\subsection{Direct Access}
\label{sec:da}

A problem that is closely related to ranked enumeration is
\emph{direct access}~\cite{carmeli23direct,carmeli20random,eldar24direct},
which asks whether it is possible to efficiently jump to arbitrary positions
in the (implicit) output array, after a preprocessing phase. 
Ranked enumeration is a special case of this problem, where the accessed positions are $1, 2, 3,\ldots$

Interestingly, the absence of disruptive trios (\Cref{def:trio}) that 
describes the feasible lexicographic orders for
the algorithm of \Cref{sec:lex}
also appears as a necessary condition for achieving direct access with quasilinear preprocessing and polylogarithmic delay~\cite{carmeli23direct} (assuming \sparseBMM{}).
The other necessary condition is for the (self-join-free) acyclic CQ
to be $L$-connex for the variables $L$ that appear in the lexicographic order;
similarly to the free-connex property,
this means that the query remains acyclic when we add a hyperedge consisting of the $L$ variables.
These two conditions are also sufficient for acyclic CQs and thus, provide a dichotomy for self-join-free CQs, under \sparseBMM{} and \hyperclique{}.

A similar, but much more restrictive on the positive side, dichotomy has been established for SUM~\cite{carmeli23direct}.
Going beyond quasilinear preprocessing time,
Bringman et al.~\cite{bringmann22da} 
derived precise bounds for each JQ and lexicographic order,
Eldar et al.~\cite{eldar24direct} considered queries with aggregation,
while Tziavelis et al.~\cite{tziavelis23quantiles}
studied the problem of a single access where no preprocessing is required.
}

\section{Conclusion and Future Outlook}
\label{sec:conclusion}

In this paper, we explored the problem of ranked enumeration without fully materializing
the query result. 
We \revchange{discussed} 
how, for acyclic join queries, certain lexicographic orders can naturally be produced by the unranked enumeration algorithm (with an additional sorting of individual relations).
However, not all orders
can be handled in this straightforward way.
With additional preprocessing and data structures for prioritization,
we presented an extended algorithm capable of handling more complex ranking functions, including SUM.
Notably, for free-connex CQs, this approach achieves 
$\TT(k)=\Otilde(n+k)$ \emph{for any subset-monotone ranking function},
and no other self-join-free CQ admits this guarantee (under common hypotheses).
Broader classes of queries are also within the reach of the algorithm,
as long as they can be efficiently reduced to a union of acyclic and free-connex CQs.

These results \revchange{are part of} an extensive line of research in database theory,
focused on the computational tasks that can be
efficiently performed on query results without explicitly materializing them. 
The goal is to offer the illusion of a materialized result, while
the actual operations are executed directly on the database. 
Beyond ranked enumeration \revchange{and direct access}, related tasks include aggregation~\cite{abo16faq}, linear regression~\cite{olteanu16record}, and $k$-means clustering~\cite{moseley21kmeans}, among others. 

One of the areas lacking a refined understanding for ranked enumeration
is the complexity landscape for ranking functions.
Although some orders are algorithmically easier to achieve than others within the subset-monotone class,
their complexity is the same, modulo logarithmic factors.
On the other end of the spectrum, for arbitrary black-box ranking functions, no strong guarantees can be achieved.
What about the space in-between?
To contrast this with
the problem of direct access, more intriguing, polynomial-time separations are known
even within the class of lexicographic and SUM ranking functions.
Mapping out properties of ranking functions and their impact on
complexity is an interesting research direction.

Similarly, more work is needed to understand the fundamental difficulty of
ranking.
For instance, are there surprising cases where ranked enumeration is harder than unranked?
One avenue to approach this question is to study CQs with ``long'' 
inequalities
(in contrast to the ``short'' inequalities of \Cref{sec:beyond_acyclic}).
For queries, such as $Q(x_1, x_2, x_3, x_4) \allowbreak\datarule\allowbreak R(x_1, x_2),\allowbreak S(x_2, x_3),\allowbreak T(x_3, x_4),\allowbreak x_1 \!<\! x_4$,
it is known that unranked enumeration can be achieved with $\Otilde(n+k)$~\cite{qichen22comparisons},
yet ranked enumeration has not been studied.
Another avenue is to consider different classes of circuits~\cite{amarilli24ranked} instead of CQs in order to find such a separation.

The relationship between ranked enumeration and top-$k$ can also lead to interesting questions.
Top-$k$ introduces two relaxations, the exact impact of which is not entirely clear: (1) $k$ is a small constant, and (2) $k$ is known in advance.

Finally, parallelization is a natural, but challenging, direction.
The prioritization of answers required by ranked enumeration implies a degree
of sequentiality in the computation, making a parallel adaptation non-obvious.
On the theoretical side, the widely used MPC model~\cite{beame17parallel} does not seem to be a good fit because of its batch-processing nature.

\introparagraph{Acknowledgements}
This work was supported in part by a grant from PricewaterhouseCoopers (PwC),
the National Institutes of Health (NIH) under award number R01 NS091421, and by
the National Science Foundation (NSF) under award numbers CAREER IIS-1762268
and IIS-1956096.
Nikolaos Tziavelis was additionally supported by a Google PhD fellowship.
Any opinions, findings, and conclusions or recommendations expressed in this paper are those of the authors and do not necessarily reflect the views of the funding agencies.

\renewcommand*{\bibfont}{\footnotesize}
\printbibliography	%

\end{document}